\documentclass[10pt,preprint]{aastex}

\newcommand{\D}[1]{\, d #1 \,}
\newcommand{\vc}[1]{\vec{#1}}
\newcommand{\avg}[1]{\left< #1 \right>}

\newcommand{\F}{\mathcal{F}}
\newcommand{\s}{\mathcal{S}}

\shorttitle{Extending Big Power Law}
\shortauthors{Chepurnov \& Lazarian}

\begin{document}

\title{Extending Big Power Law in the Sky with Turbulence Spectra from WHAM data}

\author{A. Chepurnov, A. Lazarian}
\affil{Department of Astronomy, University of Wisconsin, Madison, USA}

\begin{abstract}
We use the data of Wisconsin H$\alpha$ Mapper (WHAM) to test the hypothesis of whether the amplitudes and spectrum of density fluctuations measured by WHAM can be matched to the data obtained for interstellar scintillations and scattering. To do this, first of all, we adjusted the mean level of signal in the adjacent patches of the data. Then, assuming that the spectrum is Kolmogorov, we successfully matched the amplitudes of turbulence obtained from the WHAM data and the interstellar density fluctuations reported in the existing literature. As a result, we conclude that the existing data is consistent with the Kolmogorov cascade which spans from $10^6$ to $10^{17}$ $m$.
\end{abstract} 

\keywords{methods: data analysis --- turbulence --- ISM: lines and bands --- techniques: spectroscopic}


\section{Introduction}

The interstellar medium (ISM) is turbulent on scales ranging from AUs to kpc (see Elmegreen \& Scalo 2004 for a review), with an embedded magnetic field that influences almost all of its properties. MHD turbulence is generally accepted to be of key importance for fundamental astrophysical processes, e.g. star formation, propagation and acceleration of cosmic rays (McKee \& Ostriker 2007, Lazarian et al. 2009). 
 
A substantial insight into the scale density fluctuations has been obtained via studies of radio scattering and scintillations. It was shown in Armstrong et al. (1995) that the spectrum of density inhomogeneties is consistent with the Kolmogorov turbulence over a wide range of scales from $10^6$ $m$ to $10^{13}$ $m$ (see also Spangler \& Gwinn 1990, Wang et al. 2005, You et al. 2007). The theoretical explanation of this effect is provided within Goldreich-Sridhar (1995) model of turbulence, where Kolmogorov spectrum of magnetohydrodynamic fluctuations is expected (see also Cho \& Lazarian 2003, Beresnyak \& Lazarian 2009). But what is happening at larger scales?
 
The question was addressed by a number of recent studies. For instance, Haverkorn et al. (2008) used the Southern Galactic Plane Survey data and concluded that the injection scale of the turbulence in the spiral arms of the Galaxy should be rather small, e.g. order of a parsec, while a much larger, e.g. order of 100 $pc$, outer scale is reported for the interarm regions. This, however contradicts to the smooth power laws observed for velocity fluctuations in HI (Green 1993, Lazarian \& Pogosyan 2000) and thus stimulates a further quest of density fluctuations at large scales.

In the paper below we address this question by analyzing the Wisconsin H$\alpha$ Mapper (WHAM) data (see Madsen, Haffner \& Reynolds 2002). We check the consistency of the fluctuations of WHAM and those measured by scintillations and scattering to the Kolmogorov cascade. 

In what follows, we discuss the WHAM data in Sect. 2, expected angular spectrum of fluctuations in Sect. 3, determination of the spectrum's free parameters in Sect. 4 and provide discussions in Sect. 5. In Appendix A we describe the algorithm of correcting background level errors.

\section{WHAM data and analysis}

The Wisconsin H$\alpha$ Mapper has surveyed the distribution and kinematics of ionized gas in the Galaxy above declination $-30^o$. The WHAM Northern Sky Survey (WHAM-NSS) has an angular resolution of $1^o$ and provides the first absolutely calibrated, kinematically resolved map of the H$\alpha$ emission from the warm ionized medium (WIM) within $\pm 100$~km/s of the local standard of rest \citep{Haf03}.

We used the publicly available WHAM integrated intensity data for high Galactic latitudes. The choice is motivated by reducing the noise from localized regions, e.g. HII regions, in the galactic disk. 

The integrated WHAM map consists of patches of contiguously observed data of size 7x7 pointings. The background levels in the patches are different due to different observation conditions. Such patching causes a systematic error, which requires a proper handling, especially in the low-signal area near the North Galactic Pole (NGP).

Our data handling included the equalization of background signal level in the adjacent patches of the emission map. The particular algorithm is described in the Appendix \ref{sect:syserr}. Figure 1 shows the original and the processed data.  One may notice that the original data is patchy and one expects these borders of the patches to contribute strongly into measured spectrum. The processed data exhibits fluctuations, but do not exhibit rectangular domains of varying intensity.

According to \citet{Schl98}, dust extinction in the direction of NGP is only 3\% for H$\alpha$ and is neglected in our calculations. 

The processed data is used to estimate spectrum parameters. We do this by probing signal fluctuations at two different scales with subsequent fitting of the angular spectrum model. The latter has two free parameters, which depend on the outer scale of turbulence and amplitude of electron density spectrum. The results are tested for statistical consistency.

This approach allows us to recover information about the density spectrum despite rather small dynamical range of scales characterizing the WHAM data.

As the actual observable for H$\alpha$ data is the emission measure ($EM = \int n_e^2 \D{r}$), we needed to handle the case of quadratic emissivity to bind the expected to be Kolmogorov spectrum of $n_e$ to the angular spectrum of H$\alpha$ fluctuations. 


\section{Theoretical expectations}

We attempt to test whether the WHAM data is consistent with the hypothesis of the Kolmogorov energy cascade originating at large spatial scales. To achieve this goal we, first of all, must relate the power spectrum of underlying density and the angular spectrum of intensity fluctuations. The latter is usually measured in terms of spherical harmonics. In \S3.1 and \S3.2 we derive the necessary expressions and make use of the WHAM data in \S3.3. 

\subsection{Mapping of 3D power spectrum to the angular spectrum \label{sect:mapping} }

Let $s$ be a homogeneous isotropic random field in a 3D space. Then its mapping to the angular coordinates is given by the following expression:

\begin{equation}
S(\theta,\phi) 
  \equiv \int_0^\infty w(r) s(r,\theta,\phi) \D{r}
\end{equation}

\noindent where $w$ is a window function setting the layer thickness. The correspondent correlation function depends only on $\theta$:

\begin{equation}
C(\theta) 
  \equiv \avg{S(0,0) \cdot S^*(\theta,0)}
\end{equation}

After some algebra we can get:
\begin{equation}
C(\theta) 
  = 4\pi \int_0^\infty w(r)\D{r} \int_0^\infty w(r')\D{r'} \int_0^\infty k^2\D{k}
    F(k) \frac{\sin k\sqrt{r^2-2rr'\cos\theta+r'^2}}{k\sqrt{r^2-2rr'\cos\theta+r'^2}}
\end{equation}
\noindent where $F$ is a power spectrum of $s$.

Having calculated the angular power spectrum
\begin{equation}
C_l 
  \equiv \int_0^\pi C(\theta) P_l(\cos \theta) \sin \theta \D{\theta}
\end{equation}
\noindent we obtain:
\begin{equation} \label{eq:map}
C_l 
  = 16\pi^2 \int_0^\infty k^2 \D{k} F(k)
    \left( \int_0^\infty j_l(kr) w(r) \D{r} \right)^2
\end{equation}

If we choose $w$ in a Gaussian form with the layer thickness $R$
\begin{equation} \label{eq:wgauss}
w(r)
  = e^{-\frac{r^2}{R^2}}
\end{equation}
\noindent we can write
\begin{equation} \label{eq:mapgauss}
C_l 
  = \frac{16\pi^2}{R} \int_0^\infty u^2 \D{u} F(u/R) Q_l(u)
\end{equation}
\noindent where
\begin{equation} \label{eq:q}
Q_l(u) 
  \equiv \left( \int_0^\infty j_l(uv) e^{-v^2} \D{v} \right)^2
\end{equation}

After some algebra we can find the following asymptotic expression for the kernel $Q$:
\begin{equation} \label{eq:qappr}
Q_l(u) 
  \approx \frac{\pi}{2lu^2} e^{-2 \frac{l^2}{u^2}}
\end{equation}
which holds for $l\gtrsim 15$ uniformly over $u$.

The expressions \ref{eq:mapgauss} and \ref{eq:qappr} give us a simple mapping from 3D power spectrum $F(k)$ to the angular spectrum $C_l$.

For example, setting $F$ to a power law with spectral index $\alpha$ and outer scale $L$
\begin{equation} \label{eq:flin}
F(k) 
  = \frac{F_0}{k^\alpha} \cdot e^{-\left(\frac{2\pi}{kL}\right)^2}
\end{equation}
\noindent we can evaluate Eq.~(\ref{eq:mapgauss}) analytically:
\begin{equation} \label{eq:clin}
C_l 
  = \frac{4\pi^3\Gamma\left(\frac{\alpha-1}{2}\right)}{2^{\frac{\alpha-1}{2}}}\cdot
    \frac{F_0 R^{\alpha-1}}{
      l^\alpha\cdot\left(1+\frac{1}{2}\left(\frac{2\pi R}{l\cdot L}\right)^2\right)
      ^\frac{\alpha-1}{2}
    } 
\end{equation}

\subsection{Squared Kolmogorov field \label{sect:squared}}

As the $H\alpha$ emissivity is quadratic with respect to the underlying physical field (electron density), we have to account for the related effects.

If $\s$ is the underlying field (homogeneous and isotropic), our emissivity can be written in the following form:
\begin{equation}
s(\vc{r}) 
  = \left(\avg{\s}+\Delta \s(\vc{r})\right)^2
\end{equation}
while $\avg{\Delta \s}=0$. If $\F$ is a power spectrum of $\Delta \s$, we can write the following expression for the emissivity power spectrum $F$, assuming that $\Delta \s$ is a Gaussian field:
\begin{equation} \label{eq:qeps}
F(\vc{k}) 
  = (2\avg{\s})^2 \F(\vc{k})+2F_q(\vc{k})
\end{equation}
\noindent where
\begin{equation}
F_q(\vc{k}) 
  \equiv \int \F(\vc{k}') \F(\vc{k}-\vc{k}') \D{\vc{k}'}
\end{equation}
Let us find an approximation for $F_q$. 

Let us assume that $\F$ has Kolmogorov spectral index and a low-wavenumber cutoff at $k_0$:
\begin{equation}
\F(k) 
  = \frac{\F_0}{k^{11/3}} \cdot e^{-\frac{k_0^2}{k^2}}
\end{equation}
Then, $F_q$ can be approximated using the method of matched asymptotic expansion for the asymptotic limits $k \to 0$ and $k \to \infty$. The asymptotic $k \to \infty$ should decrease as $k^{-11/3}$ with a scaling factor of $2\int\F(\vc{k})\D{\vc{k}}$, which can can be illustrated as follows. As the spectral index is greater than 3, the bulk of the integrated value of $\F$ exists in a region near $k = 0$. For very large $k$, the auto-convolution consists of two locations of large integrated value, i.e. around $k' = 0$ and $k'-k = 0$, i.e. separated by a distance $k$. Interpolating with the correct behavior in between these two asymptotic values allows $F_q$ to be approximated as\footnote{See Fig. \ref{fig:fq} for comparison of directly calculated $F_q$ with its approximation}:
\begin{equation}
F_q(k) 
  \approx \F_0^2 \frac{J_0}{\left(\left(\frac{J_0}{2J_1}\right)^{6/11}k^2+1\right)^{11/6}}
\end{equation}
\noindent where
\begin{equation}
J_0
  \equiv \frac{4\pi}{\F_0^2} \int_0^\infty \F^2(k)k^2\D{k}
\end{equation}
\begin{equation}
J_1
  \equiv \frac{4\pi}{\F_0} \int_0^\infty \F(k)k^2\D{k}
\end{equation}
Evaluating $J_0$ and $J_1$ we can write the following expression for $F_q$:
\begin{equation}
F_q(k) 
  \approx \frac{F_{q0}}{\left(a\frac{k^2}{k_0^2}+1\right)^{11/6}}
\end{equation}
\noindent where
\begin{equation}
F_{q0} = \frac{1.514\F_0^2}{k_0^{13/3}}, \;\; a=0.1842
\end{equation}
Applying Eq.~(\ref{eq:mapgauss}) to $F_q$, after some algebra we get
\begin{equation} \label{eq:cquad}
C_{q,l} 
  = 1.48\cdot 10^3\cdot \frac{\F_0^2 R^{8/3}}{\left(\frac{2\pi}{L}\right)^{2/3}}\cdot
    \frac{1+3.21\cdot e^{-1.43\frac{lL}{2\pi R}}}{
      l^{11/3}\cdot\left(1+5.0\left(\frac{2\pi R}{l\cdot L}\right)^2\right)
      ^{4/3}
    } 
\end{equation}

Finally, using Eq.~(\ref{eq:clin}) for the first term in Eq.~(\ref{eq:mapgauss}) when $F$ is given by Eq.~(\ref{eq:qeps}), we have the following expression for the angular spectrum:

\begin{equation} \label{eq:kolm_sq}
\begin{array}{ll}
C_l
  &= 1.72 \cdot 10^2 \cdot \frac{\avg{\s}^2 \F_0 R^{8/3}}{l^{11/3}(1+0.5(2\pi R)^2/(l L)^2)^{4/3}}\\
  &+ 2.96 \cdot 10^3 \cdot \frac{\F_0^2 R^{8/3}}{(2\pi/L)^{2/3}} 
     \cdot \frac{1+3.21e^{-1.43\frac{l L}{2\pi R}}}{l^{11/3}(1+5.0 (2\pi R)^2/(l L)^2)^{4/3}}
\end{array}
\end{equation}

\subsection{Expected spectrum of fluctuations}

In this section we calculate the expectations for the angular spectrum of H$\alpha$ fluctuations assuming that the emissivity is proportional to the squared random field having Kolmogorov spectrum (i.e. electron density). We shall start from Eq.~(\ref{eq:kolm_sq}), accounting for the following expression of H$\alpha$ intensity\footnote{Electron temperature of 8000K and $\gamma = 0.9$ is assumed, see \citet{Arm95}} (Smoot 1998):
\begin{equation} \label{eq:intens}
\begin{array}{ll}
I
  & = 0.36 \cdot \frac{EM}{cm^{-6} pc} \left(\frac{T}{10^4 K} \right)^{\gamma} \\
  & = 0.29 \cdot \int_{los} n_e^2 dr
\end{array}
\end{equation}
\noindent where electron density $n_e$ is measured in $cm^{-3}$, $r$ in $pc$ and $I$ in Rayleighs. 

Assuming that $\avg{n_e}$ is the mean electron density, $R$ is the scale-height of the $H_\alpha$ layer, $L$ is the injection scale of the turbulence and $F_0$ is the amplitude of the electron density spectrum one can write for the spherical harmonics:
\begin{equation} \label{3.1}
\begin{array}{ll}
C_l
  &= 1.42 \cdot \frac{\avg{n_e}^2 F_0 R^{8/3}}{l^{11/3}(1+0.5(2\pi R)^2/(l L)^2)^{4/3}}\\
  &+ 2.41 \cdot \frac{F_0^2 R^{8/3}}{(2\pi/L)^{2/3}} 
     \cdot \frac{1+3.21e^{-1.43\frac{l L}{2\pi R}}}{l^{11/3}(1+5.0 (2\pi R)^2/(l L)^2)^{4/3}}
\end{array}
\end{equation}
where $\avg{n_e}$ is measured in $cm^{-3}$, $F_0$ -- in $m^{-20/3}$, $R$ and $L$ -- in $pc$.

Let us rewrite Eq.~(\ref{3.1}) as follows:

\begin{equation} \label{3.3}
\begin{array}{ll}
C_l
   &=1.42 \cdot p q^{5/3} \alpha^{8/3} \cdot \frac{\avg{n_e}^2 R_0}{l^{11/3}(1+0.5(2\pi \alpha q/l)^2)^{4/3}} \\
   &+0.709 \cdot p^2 q^{8/3} \alpha^{8/3} \cdot \frac{1+3.21e^{-1.43 l/(2\pi \alpha q)}}{l^{11/3}(1+5.0 (2\pi\alpha q/l)^2)^{4/3}}
\end{array}
\end{equation}

where $p=F_0 L^{5/3}$, $q=R_0/L$ and $\alpha=R/R_0$. For the polar region we can assume $\alpha\approx 1$. 

One can see that the the angular spectrum given by Eq.~(\ref{3.3}) has two free parameters, $p$ and $q$, which can be evaluated from the observational data. Other parameters are taken as follows: $\avg{n_e}=0.010 cm^{-3}$ \citep{Cor02}, $R_0 = 1800$ $pc$ \citep{Gae08}.

The range of scales that is sampled with the WHAM is rather short. Therefore we shall make our comparison between the data and the model using a measure of the dispersion of the signal $s(\vc{\rho}')$ over a circular area $\Omega_a (\vc{\rho})$ of radius $a$ and centered at $\vc{\rho}$:
\begin{equation} \label{3.4}
D(\vc{\rho})
  =\frac{1}{\pi a^2} \int_{\Omega_a(\vc{\rho})}
   \left(s(\vc{\rho'})-\frac{1}{\pi a^2}\int_{\Omega_a(\vc{\rho})} s(\vc{\rho''})\D{\vc{\rho''}}\right)^2 \D{\vc{\rho'}}
\end{equation}

When applied to the angular spectrum given by Eq.~(\ref{3.3}), it provides
\begin{equation} 
D
  =\frac{1}{(2\pi)^2} \int \Phi(a, \kappa)C(\kappa) \D{\vc{\kappa}} = pf_1(a, \alpha, q) + p^2 f_2(a, \alpha, q)
\label{3.5}
\end{equation}
where
\begin{equation} \label{3.6}
\Phi(a, \kappa)=(1-\phi(a,\kappa))\cdot \phi(r, \kappa)
\end{equation}
and $\phi(a,\kappa)=\left(\frac{2}{a \kappa} J_1(a \kappa)\right)^2$ and $r$ is the pixel radius, which in the case of WHAM data is $0.5^o$. The functions $f_i$ in Eq.~(\ref{3.5}) are given by  
\begin{equation} \label{3.7}
 \begin{array}{ll}
& f_1(a,\alpha, q)=0.256 \frac{1}{2\pi} 
  \int^{\infty}_{2}\frac{\alpha^{8/3} q^{5/3}}{l^{11/3}(1+0.5(2\pi \alpha q/l)^2)^{4/3}} \Phi(a,\kappa) \kappa \D{\kappa} \\
& f_2(a,\alpha,q)=0.709 \frac{1}{2\pi} 
  \int^{\infty}_{2} \frac{ \alpha^{8/3} q^{8/3}(1+3.21e^{-1.43 l/(2\pi \alpha q)})}{l^{11/3}(1+5.0 (2\pi\alpha q/l)^2)^{4/3}} \Phi(a,\kappa) \kappa \D{\kappa}\\
\end{array}
\end{equation}
In what follows we use Eq.~(\ref{3.5}) to obtain the dispersions for different values of $a$.

\section{Probability distributions of $p$ and $q$}


For the analysis below we used two sets of maps with radii\footnote{This numbers are taken to get maximal dynamical range from the available data} $a_1=2.27$ degrees and $a_2=8.33$ degrees. For those values of $a$ the dispersions calculated using the WHAM data are $D_1=7.36\cdot 10^{-3}\pm 0.72\cdot 10^{-3}$ and $D_2=10.42\cdot 10^{-3}\pm 1.05\cdot 10^{-3}$, where the instrumental error of 0.04 R is accounted for.

To estimate the consistency of our assumption of the Kolmogorov turbulence present in warm gas sampled by WHAM we calculate the probability distribution of the free parameters $p$ and $q$. We start from the mutual probability of $D_1$ and $D_2$ roughly assuming that these values are independent and have normal PDF: 
\begin{equation}  \label{4.1}
P_D(D_1, D_2)
  =\frac{1}{2\pi \sigma_{D1} \sigma_{D2}}\exp\left(-\frac{(D_1-\bar{D_1})^2}{2\sigma^2_{D_1}}-\frac{(D_2-\bar{D_2})^2}{2\sigma^2_{D_2}}\right)
\end{equation}

Solving Eq.~(\ref{3.5}) in terms of $p$, one gets
\begin{equation} \label{4.2}
 p
  =F(a_i, \alpha_i, q, D_i), \; i=1,2
\end{equation}
\noindent where
\begin{equation}
F(a_i, \alpha_i, q, D_i)
  = \frac{(f_1(a_i,\alpha_i,q) + 4 f_2(a_i,\alpha_i,q)\cdot D_i)^{1/2} -f_1(a_i, \alpha_i, q)}{2f_2(a_i, \alpha_i, q)}
\label{4.3}
\end{equation}
 
Below we shall use Eq.~(\ref{4.2}) to derive the probability distribution $P(q, p)$ that will be used to estimate the parameters $p$ and $q$ and their accuracy. If we write the elementary probability in the ${(D_1, D_2)}$ space as $dP=P_D(D_1, D_2) dD_1 dD_2$, we get the following Jacobian of the transformation:
\begin{equation}
dS
  =\frac{\frac{\partial}{\partial D_1} F(a_1, \alpha_1, q, D_1) \frac{\partial}{\partial D_2} F(a_2, \alpha_2, q, D_2)}
        {\left| \frac{\partial}{\partial q} F(a_1, \alpha_1, q, D_1) - \frac{\partial}{\partial q} F(a_2, \alpha_2, q, D_2)\right|} 
   \cdot dD_1 dD_2
\label{4.4}
\end{equation}
which results in the expression for the probability:
\begin{equation}
P(p,q)
  = \frac{\left| \frac{\partial}{\partial q} F(a_1, \alpha_1, q, D_1) - \frac{\partial}{\partial q} F(a_2, \alpha_2, q, D_2)\right|}
         {\frac{\partial}{\partial D_1} F(a_1, \alpha_1, q, D_1) \frac{\partial}{\partial D_2} F(a_2, \alpha_2, q, D_2)} 
    \cdot P_D(D_1,D_2)
 \label{4.5}
 \end{equation}
where one should substitute $D_i$ for its expression, i.e. $D_i=p f_1(a_i, \alpha_i, q)+ p^2 f_2(a_i, \alpha_i, q)$.
 
The practical implementation of the finding probability distributions $P(p)$ and $P(q)$ is illustrated in Figures 2 and 3.  The both curves were re-normalized. Initially they had encompassed the area of 0.85, which meant that the model spectrum was inconsistent with the observational data with the probability of 0.15. Using the calculated PDFs, we finally have:
\begin{equation}
q = R_0/L = 20^{+22}_{-8},\;\;\; p = F_0 L^{5/3} = 0.54^{+0.14}_{-0.10}
\end{equation}
\noindent for confidence probability 0.68.
Setting $R_0 = 1800pc$ \citep{Gae08}, we can find the injection scale $L$ and electron density spectrum amplitude $F_0$:
\begin{equation}
L = 90^{+60}_{-50} pc
\end{equation}
\begin{equation}
F_0 = 3^{+9}_{-2}\cdot 10^{-4} m^{-20/3}
\end{equation}
The related spectrum is shown on the Figure \ref{fig:bpl}.

\section{Discussion}

The Big Power Law in the Sky with the spectral index coincided with the one in Kolmogorov turbulence was discussed in the paper of Armstrong et al. (1995). The reliable data used in their plot reflect the density fluctuations at scales from $10^6$ to $10^{13}$ $m$ measured via scintillation and the electron scattering technique.  For larger astrophysical scales the plot in Armstrong et al. (1995) contains either upper limits or unreliable data, for instance, rough estimates based on turbulent velocity. 

This paper contains the first piece of evidence that the spectrum of density fluctuations at very small scales shown in Armstrong et al. (1995) agrees well with the spectrum of the density fluctuations measured at scales of $10^{17}$ m, if we assume that the scale height of the free-free emitting layer is $1800pc$. The found spectrum amplitude $F_0 = 3 \cdot 10^{-4} m^{-20/3}$ is within the error bar from the value $3.2 \cdot 10^{-4} m^{-20/3}$ found by \citet{Cor91} from pulsar scintillations. This is a remarkable extension of the Big Power Law in the Sky. 

This is suggestive of the energy being injected at scales of $40 \div 150 pc$ in the Galaxy and cascading up to very small scales. Below we discuss whether this case is a plausible one.

Does this picture of the large scale turbulent energy cascade look reasonable? Density information alone cannot answer this question. Turbulence is a dynamical process, in which density can be used only as a tracer. More direct information is available through velocity studies. Our estimation is also in agreement with the expected value of 100 pc associated with supernova explosions (see for instance \citet{Hav08}). Studies in Chepurnov et al. (2009) of the velocity turbulence using the Velocity Coordinate Spectrum (VCS) provide a good fitting of the turbulence model when the injection scale is taken to be $140\pm 80$ $pc$ (VCS uses Fourier-transformed over velocity coordinate spectral data to get analytically predictable data measure, see \citet{LP00}.). In addition, MHD simulations in Cho \& Lazarian (2003) indicate that for subsonic turbulence the density spectrum follows well the velocity spectrum, which is Kolmogorov. This is however is not true for supersonic MHD turbulence, for which, according to Beresnyak, Lazarian \& Cho (2005) the density spectrum gets shallow.

We feel that the issue of the spectral slope does require further studies. With our limited dynamical range we could test the consistency of the spectral index to the Kolmogorov one. Discontinuous structures, e.g. ionized ridges of clouds make the spectrum more shallow. 

What is the Mach number of the free-free emitting layer in our Galaxy? This question was addressed by Hill et al. (2008), who compared the PDFs obtained via MHD simulations in Kowal, Lazarian \& Beresnyak (2007) and the PDFs of WHAM data. As a result a conclusion that the sonic Mach number of turbulence in free-free emitting layer, which is also frequently called the Reynolds layer, is around of 2. This is close to the subsonic range and therefore we do not expect to see substantial deviations from the Kolmogorov scaling for the random density.

All in all, the arguments above are consistent with the idea that the large scale turbulence in the Reynolds layer and small scale turbulence constituting the Big Power Law in Armstrong et al. (1995) are the parts of the universal turbulence cascade. This is also consistent with other arguments, for instance, with the theoretical arguments on the energy injection in turbulence. Both of the leading ideas on the injection of turbulent energy, i.e. via supernovae explosions and via the magnetorotational instability, inject energy at large scale, e.g. larger than 30 pc. This energy is bound to create a cascade in high Reynolds number interstellar medium. Therefore the emergence of the extended turbulent cascade is expected.  

The present study shows the consistency of the data with the Kolmogorov cascade in interstellar gas spanning over 10 decades. Further studies combining various data sets, including those of velocity and magnetic field will clarify the nature of the turbulence cascade in the Galaxy.

\section{Acknowledgments}

Authors thank the WHAM team for their great work and making the data publically available. The Wisconsin H-Alpha Mapper is funded by the National Science Foundation. 

We acknowledge the NSF grant AST 0808118 and the support from the NSF funded Center for Magnetic Self-Organization.

Authors thank the anonymous referee for the useful comments which improved our paper substantially.

\appendix

\section{Removing the systematic error \label{sect:syserr}}

The column density WHAM map suffers from systematic error, related to uncertainty of zero levels in different observation blocks. It can be corrected by introducing artificial constant shifts for each observation block, which are set by minimization of signal differences on block borders.

Let use denote the initial signal level difference between $i$-th and $j$-th block by $c_{ij}$, if they have a common border\footnote{otherwise $c_{ij}$ is undefined} ($c_{ij}=-c_{ji}$), correction shift to be found for the $i$-th block by $d_i$, and weight for adjusting of involvement of the border between $i$-th and $j$-th blocks by $1/w_{ij}$ ($w_{ij}=w_{ji}$). We also define summation sets as $\Omega \equiv \{(i,j)| \exists c_{ij}\}$ and $\Omega_i \equiv \{j | \exists c_{ij}\}$.

With these definitions a function to minimize is as follows:
\begin{equation}
L
  = \sum_\Omega \frac{1}{w_{ij}^2}(c_{ij}+d_i-d_j)^2
\end{equation}
Taking derivative over $d_n$, we have at the minimum:
\begin{equation}
\frac{\partial L}{\partial d_n}
  = \sum_{\Omega_n} \frac{4}{w_{nj}^2}(c_{nj}+d_n-d_j)
  = 0
\end{equation}
\noindent what gives us a set of linear equations for $d_i$:
\begin{equation}
d_n \sum_{\Omega_n} \frac{1}{w_{nj}^2} - \sum_{\Omega_n} \frac{1}{w_{nj}^2} d_j 
  = -\sum_{\Omega_n} \frac{c_{nj}}{w_{nj}^2}
\end{equation}

To make these equations linearly independent we need to set one of $d_i$'s to some pre-defined value or set some other condition making the total base level of the map definite.

In our calculations we take $w_{ij}$ proportional to the mean signal over the involved blocks to correct the algorithm's tendency to solve the problems of high-magnitude blocks at a cost of distortion of low-signal areas. 

Now we shall consider the calculation of $c_{ij}$. Let us denote a set of points, that belongs to the $i$-th block by $A$ and set of points, that belongs to the $j$-th block by $B$ (we take only points in a small enough vicinity of some point on the border, so that such set of vicinities will give us a correspondent set of estimations of $c_{ij}$). The signal $y(\vc{r})$ is defined on $A \cup B$, while its values on the sets $A$ and $B$ differ by a constant. With exception of this difference, we consider $y(\vc{r})$ to be continuous and smooth enough.

To estimate this difference on a discrete grid $\{\vc{r}_i\}$ we shall approximate $y$ using a polynomial functional basis $\{f_n(\vc{r})\}$, $n=1...N$, with account for the fact that the constant term is different in $A$ and $B$ (having the values $c_A$ and $c_B$ respectively). The other expansion coefficients $c_n$ are equal for the both sets.

Having applied the method of least squares, we obtain the following system of $N+2$ equations for $c_n$, $c_A$ and $c_B$:
\begin{equation}
\left(
\begin{array}{lll}
\sum_{A\cup B} f_n(\vc{r}_i)f_m(\vc{r}_i) & \sum_A f_m(\vc{r}_i) & \sum_B f_m(\vc{r}_i) \\
\\
\sum_A f_n(\vc{r}_i) & \sum_A 1 & 0 \\
\\
\sum_B f_n(\vc{r}_i) & 0 & \sum_B 1 \\
\end{array}
\left|
\left|
\begin{array}{lll}
\sum_{A\cup B} y_i f_m(\vc{r}_i) \\
\\
\sum_A y_i \\
\\
\sum_B y_i \\
\end{array}
\right.
\right.
\right)
\end{equation}

Having solved it, we take $c_A-c_B$ as an estimation for $c_{ij}$. For the final value of $c_{ij}$ we take an average of different estimations, calculated along the border.

\begin{figure}
\begin{center}
\plottwo{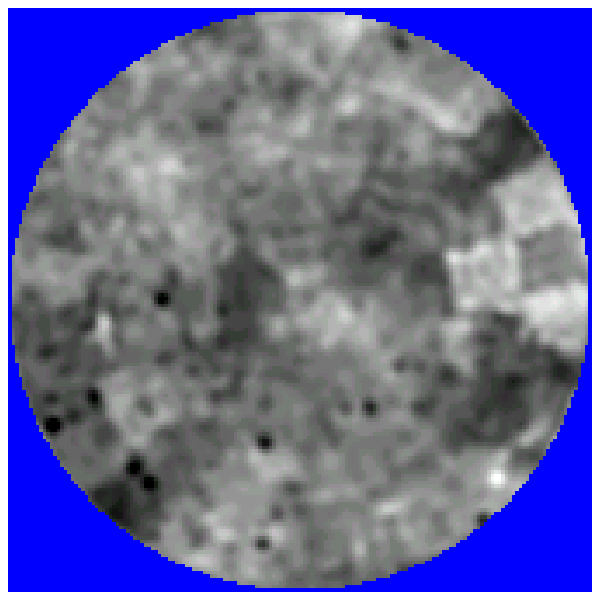}{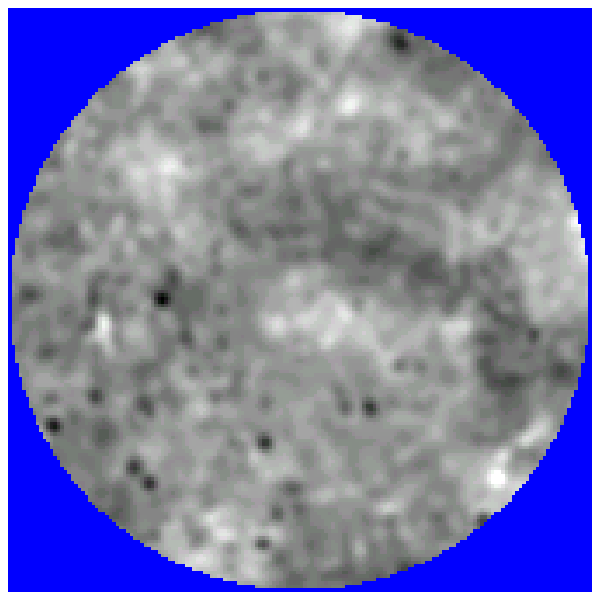}
\end{center}
\caption{The part of WHAM map near NGP (the center of map), raw and cleaned. Map radius is $25^o$. Longitude $l=0$ points to the down side of the map.  \label{fig:data}} 
\end{figure}

\begin{figure}
\begin{center}
\plotone{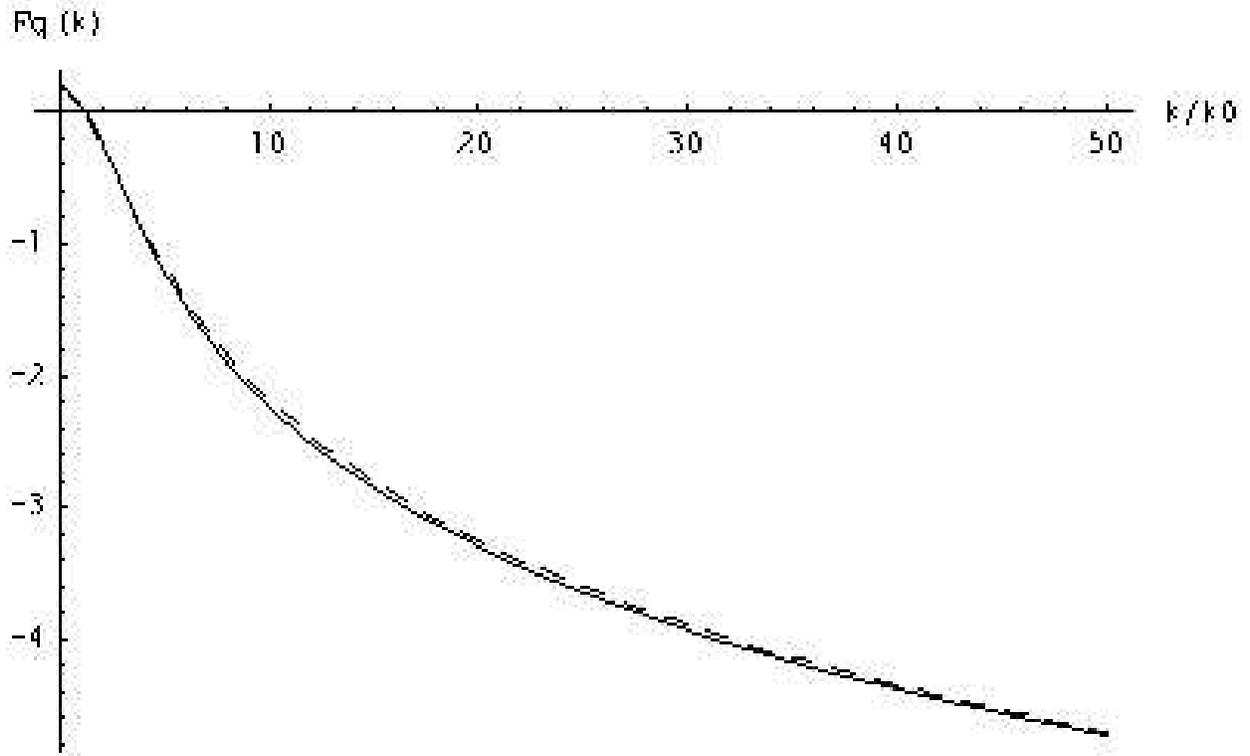}
\end{center}
\caption{$F_q$, direct calculation (solid line) and approximation (dashed line)\label{fig:fq}} 
\end{figure}

\begin{figure}
\begin{center}
\plotone{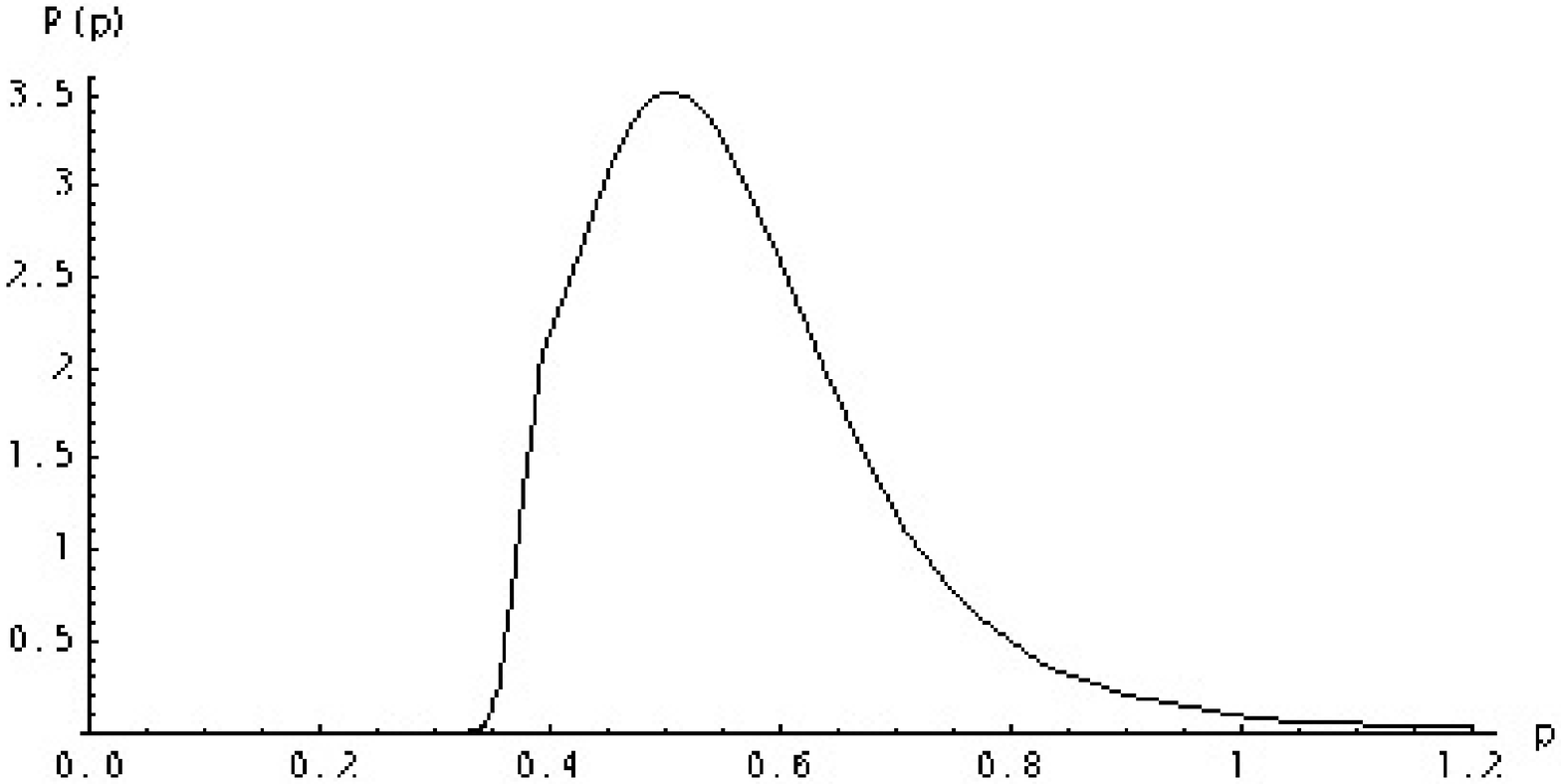}
\end{center}
\caption{The probability distribution function of the spectrum parameter $p$\label{fig:p}} 
\end{figure}

\begin{figure}
\begin{center}
\plotone{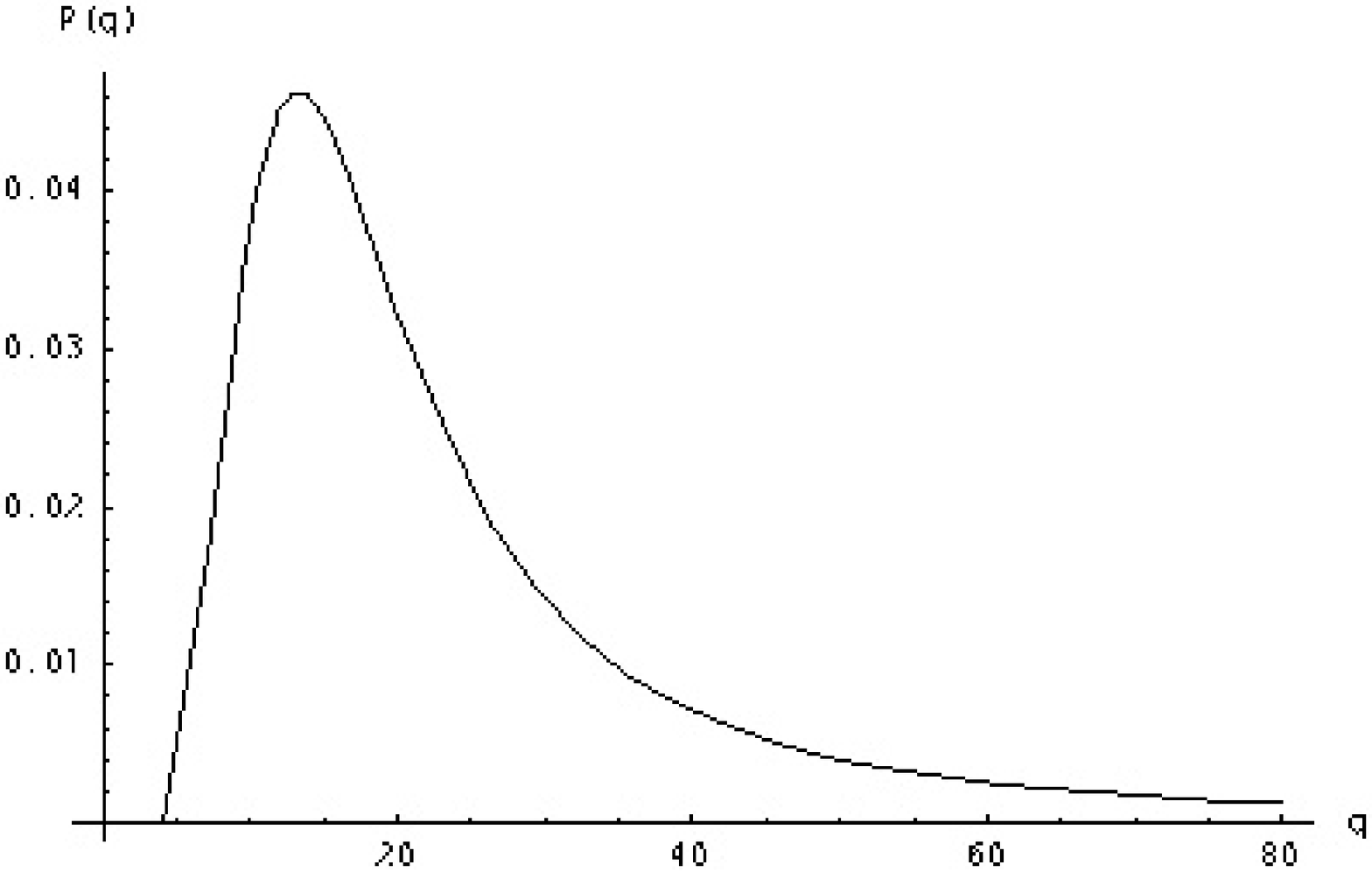}
\end{center}
\caption{The probability distribution function of the spectrum parameter $q$\label{fig:q}} 
\end{figure}

\begin{figure}
\begin{center}
\plotone{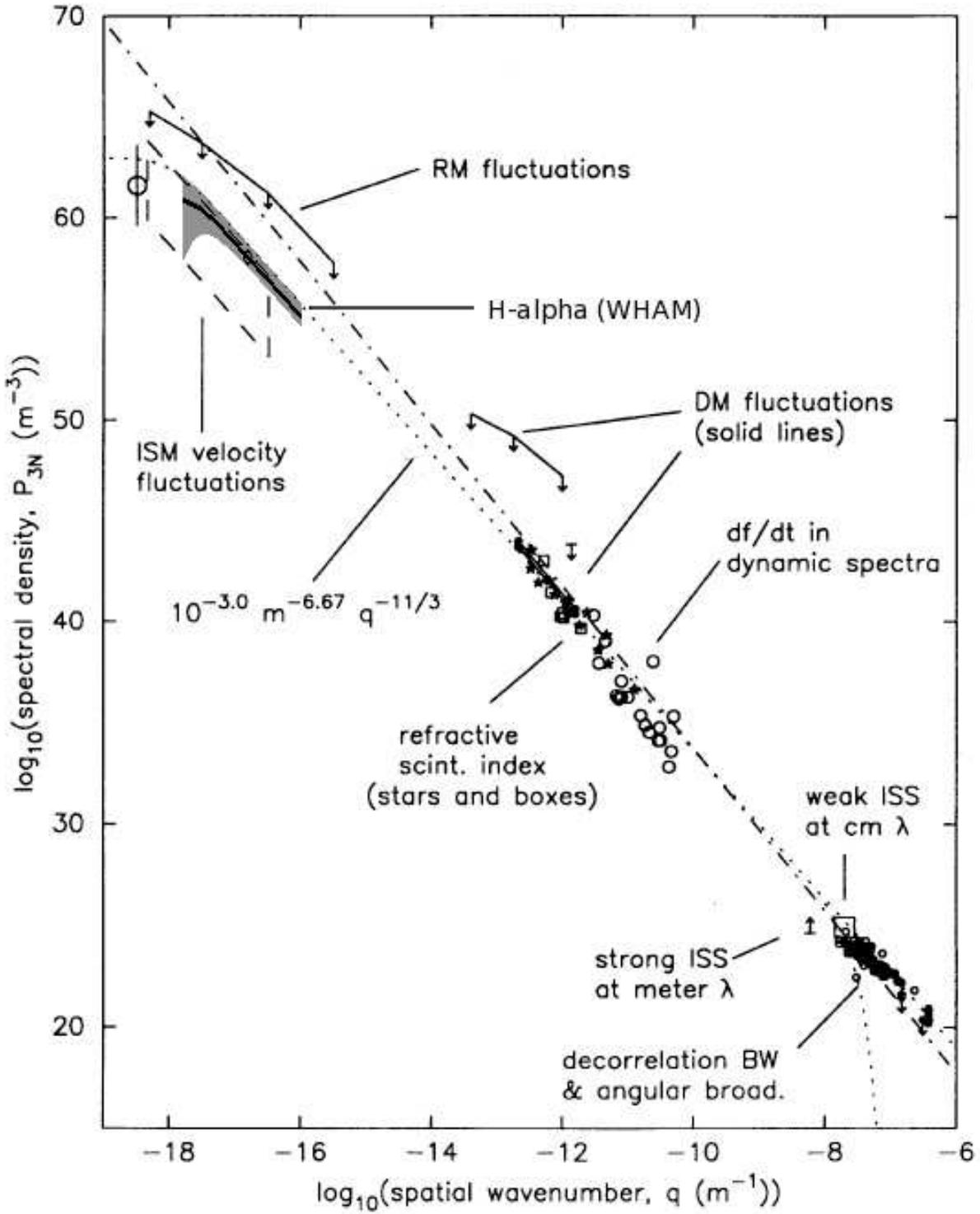}
\end{center}
\caption{WHAM estimation for electron density overplotted on the figure of the Big Power Law in the sky figure from Armstrong et al. (1995). The range of statistical errors is marked with the gray color.\label{fig:bpl}} 
\end{figure}

\end{document}